\documentclass[letterpaper]{article} 
\usepackage {aaai25}  
\usepackage{times}  
\usepackage{helvet}  
\usepackage{courier}  
\usepackage[hyphens]{url}  
\usepackage{graphicx} 
\usepackage{natbib}  
\usepackage{caption} 
\usepackage{algorithm}
\usepackage{algorithmic}

\usepackage{newfloat}
\usepackage{listings}
\DeclareCaptionStyle{ruled}{labelfont=normalfont,labelsep=colon,strut=off} 
\floatstyle{ruled}
\newfloat{listing}{tb}{lst}{}
\floatname{listing}{Listing}
\title{Measuring What Matters: Connecting AI Ethics Evaluations to System Attributes, Hazards, and Harms }
\author {
    Shalaleh Rismani \textsuperscript{\rm 1},
    Renee Shelby \equalcontrib\textsuperscript{\rm 2},
    Leah Davis \equalcontrib\textsuperscript{\rm 1}, 
    Negar Rostamzadeh \textsuperscript{\rm 2}, 
    AJung Moon \textsuperscript{\rm 1}
}
\affiliations {
    \textsuperscript{\rm 1} McGill University\\
    \textsuperscript{\rm 2} Google Research\\
}

\usepackage{bibentry}

\begin{document}

\maketitle

\begin{abstract}
Over the past decade, an ecosystem of measures has emerged to evaluate the social and ethical implications of AI systems, largely shaped by high-level ethics principles. These measures are developed and used in fragmented ways, without adequate attention to how they are situated in AI systems. In this paper, we examine how existing measures used in the computing literature map to AI system components, attributes, hazards, and harms. Our analysis draws on a scoping review resulting in nearly 800 measures corresponding to 11 AI ethics principles. We find that most measures focus on four principles -- fairness, transparency, privacy, and trust -- and primarily assess model or output system components. Few measures account for interactions across system elements, and only a narrow set of hazards is typically considered for each harm type. Many measures are disconnected from where harm is experienced and lack guidance for setting meaningful thresholds. These patterns reveal how current evaluation practices remain fragmented, measuring in pieces rather than capturing how harms emerge across systems. Framing measures with respect to system attributes, hazards, and harms can strengthen regulatory oversight, support actionable practices in industry, and ground future research in systems-level understanding.
\end{abstract}

%

\section{Introduction}

\begin{figure*}[t]
  \centering
  \includegraphics[width=\textwidth]{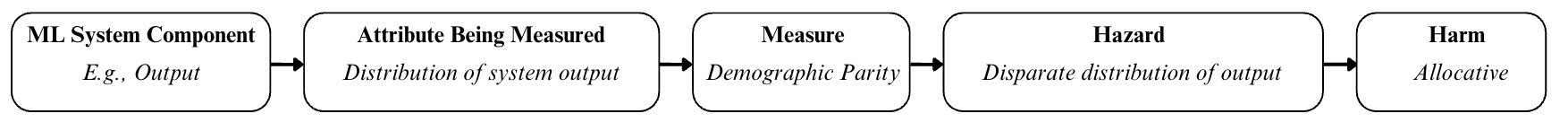}
  \caption{Example of the relational sequencing between a system component, its attribute, measure, hazard, and harm.}
  \label{fig:analyticalframe}
\end{figure*}

How do we know whether an Artificial Intelligence (AI) system adheres to AI ethics principles, and how does such adherence relate to the evaluation of algorithmic harm? As AI systems become more deeply embedded in high-stakes domains, there is a growing need to go beyond technical performance and examine the adverse impact these systems have on individuals and communities. This need is reflected in recent regulatory and policy initiatives, including the European Union’s (EU) AI Act and the National Institute of Standards and Technology's (NIST) AI Risk Management Framework \cite{EUAIAct,NISTRMF}. These initiatives emphasize establishing appropriate measurement protocols to identify and mitigate sociotechnical harm, define thresholds for unacceptable system behavior, and enable ongoing oversight and monitoring \cite[Article 9] {EUAIAct-art9}. These regulatory advances are supported by, and often built directly on, the wide range of approaches developed by the AI ethics and responsible AI (RAI) communities to assess normative qualities of AI systems (e.g., fairness \cite{Morley2023-nj}).

%
Despite the proliferation of measures designed to assess AI ethics principles, current evaluation practices fall short in two key ways. First, from a measurement theory standpoint, many existing measures lack construct validity and reliability-- they do not consistently or accurately capture the normative qualities they aim to assess  \cite{Jacobs2021-bj, Hutchinson2022-rf}. 
Without such guarantees, these measures risk producing misleading or incomplete insights about an AI system’s attributes and performance \cite{Jacobs2021-ts, Xiao2023-aa,Salaudeen2025-ph}. 

Second, most measures focus on isolated components-- such as models or datasets-- rather than the AI system as a whole or cross-component interactions. Yet, evaluating an individual component is not equivalent to assessing a systems-level property \cite{Barocas2021-sr,Vera_Liao2023-gv}. This is especially problematic from a system safety perspective, where safety is understood as an \textit{emergent} property arising from the interactions among technical, human, and organizational components \cite{Leveson2011-fo,Dobbe2022-ql}. Even when each component functions as intended, harms can emerge from inappropriate interactions.

In this paper, we build on this second criticism to critically examine the efficacy of measures used to assess an AI system's adherence to AI ethics principles, hereby referred to as RAI measures. In particular, we reflect on the role that these measures play in identifying, preventing, and responding to sociotechnical harms stemming from complex systems.
We address the following research questions:

\begin{itemize}
    \item \textbf{RQ1:} How do measures of adherence to AI ethics principles map to different components of an AI system, and what attributes do they capture? 
    \item \textbf{RQ2:} What types of hazards and harms do these measures signal? 

\end{itemize}

We conducted a scoping review of computing literature that claims to measure adherence to AI ethics principles. We extracted a total of 791 measures from 257 academic articles and performed a reflexive analysis to examine how these measures relate to system components, attributes,  hazards, and harms.
We find that existing measures are unevenly distributed across principles. Fairness, transparency, privacy, and trust account for the vast majority of measures, while principles like dignity, responsibility, and sustainability remain underrepresented. Most measures assess properties of the model and output as individual components, rather than evaluating hazards that emerge from interactions across the AI system. There is little documentation on how the measures relate to harms experienced by users, and what thresholds should warrant changes in design or deployment decisions.
Based on these findings, we call on the AI community to take on more
temporally responsive, systems-level approaches to evaluating AI systems. 

We contribute a multi-dimensional analysis of RAI measures that identifies which ethics principles, system components, and types of harm are prioritized in the literature, as well as key attributes and hazards evaluated for each harm type. Drawing on a system safety perspective, we distill cross-cutting themes that highlight gaps and opportunities for more comprehensive, integrated, and system-aware evaluation practices. 
We also provide our dataset of measures, categorized by system components, attributes, hazards, and harms, and an interactive visualization to support further exploration and engagement. These contributions advance a systems-level perspective on RAI evaluation, offering concrete implications for computing researchers, industry practitioners, and policymakers. 

\section{Background}

\begin{figure*}[ht]
  \centering
  \includegraphics[width=\textwidth]{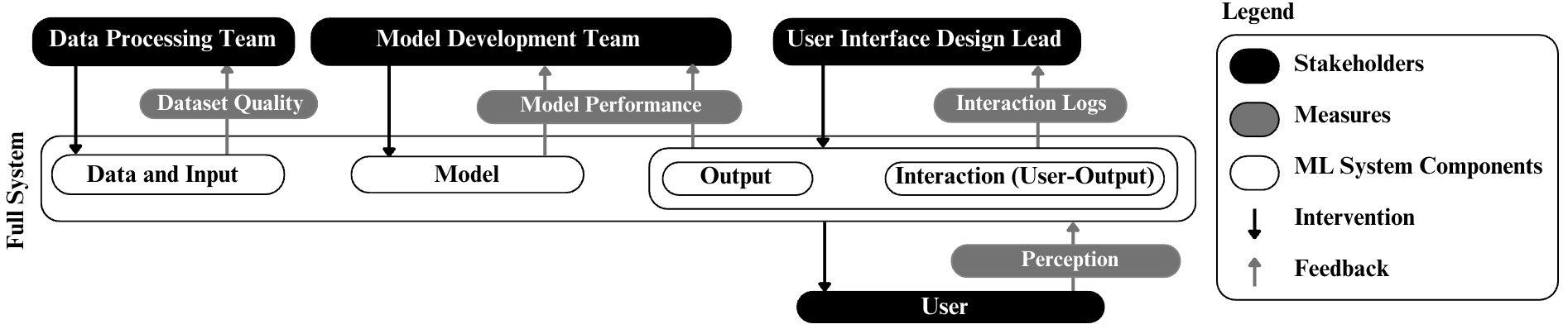}
  \caption{A conceptual model of feedback loops in a sociotechnical AI system. Measures provide signals about the status of system components, enabling stakeholders to make appropriate interventions.}
  \label{fig:controldiagram}
\end{figure*}

Measurement theory defines a measurement as the process of assigning a number to a single aspect of an object \cite{Hand2004-lp}. 
As such, a measurement is often a simplified indicator (e.g., height in centimeters) for a more complex attribute (e.g., a person's size).
In AI system development, various measurements are taken across stages like dataset curation and model training to guide refinement or improvement. In this paper, we focus on measures used by the computing community to assess the ethical normative qualities of AI systems. Next, we briefly review RAI evaluation practices and challenges, and then introduce concepts from system safety that can inform these approaches.

\subsection{RAI Evaluation Practices and Challenges}
Traditional performance measures and benchmarks have long been used to assess the quality of AI systems and track progress \cite{Reuel2024-wn,Orr2024-zg}.
In response to growing societal concerns, the scope of AI evaluation has expanded beyond performance to include social and ethical impacts since the early 2010s \cite{barocas-hardt-narayanan}.
These newer RAI assessment approaches fall into two broad categories. 

The first category includes largely qualitative assessments of impact, harm, and risk assessments designed to examine an AI system's effect on the stakeholders within its deployment context \cite{Sloane2023-wx,Watkins2021-wc}.
These methods aim to produce a context-specific understanding of harms and risks, to support governance initiatives and responsible development. 
The second category includes quantitative measures that aim to signal an AI system's normative qualities beyond simple performance metrics.
Many of these measures were developed to operationalize AI ethics principles (e.g.,  fairness, privacy, and transparency) into practice \cite{Jobin2019-kt,Morley2023-nj}.
Examples include user fairness affinity, self-reported trust, privacy budget, and embodied emissions \cite{Franklin2022-cn,Vereschak2021-gn,Carvalho2019-vh}.
Such measures may be used in risk assessments, post-deployment audits, or to evaluate trade-offs between technical performance and ethical considerations \cite{Gardner2022,Zhao2020-oh}. 

While these quantitative RAI efforts represent important progress, scholars have highlighted key limitations in their design and application. 
A central concern is that measures lack construct validity and reliability.
For instance, Jacobs and Wallach (\citeyear{Jacobs2021-bj}) use measurement theory to show how common fairness metrics often conflate theoretical constructs with their operationalizations. 
Xiao et al. (\citeyear{Xiao2023-aa}) and Blodgett et al. (\citeyear{Blodgett2021-fg}) apply similar critiques to natural language generation models and fairness benchmark datasets, respectively, demonstrating that their metrics frequently fail to capture the concepts they aim to assess.

Furthermore, most existing measures have a narrow scope, focusing on isolated components, such as model outputs, rather than evaluating AI systems in context. 
Recent work emphasizes that meaningful evaluation requires a systems-level perspective that accounts for sources of harm across AI system components, and considers for whom the system’s effects are most consequential \cite{Diaz2025-ps,Weidinger2023-pe}. Barocas et al. (\citeyear{Barocas2021-sr}) distinguish between evaluating system-wide performance to reveal deployment-related harms and assessing individual components to identify their underlying causes.  
Likewise, Liao and Xiao (\citeyear{Vera_Liao2023-gv}) critique current evaluation practices for failing to bridge the “sociotechnical gap” between model behavior and downstream human impact. 
Without evaluating harms across components and contexts, we risk a false sense of systems-level oversight. 
Lastly, measurement practices often remain ad hoc and fragmented, driven more by convenience, precedent, or tool availability rather than by systematic reasoning \cite{Rismani2023-im,Madaio2022-vt}. These challenges underscore the need for a more structured, systems-oriented approach to evaluation. 


\subsection{A System Safety Approach to Evaluation }

Recent work suggests that system safety engineering frameworks can help developers identify and mitigate potential sources of sociotechnical harms by examining the interconnected network of AI system components and actors \cite{Qi2023-zt,Dobbe2024-tk, Rismani2023-nt}. These works emphasize the vital role measurement plays in enabling safe design \cite{Dahlgren-Lindstrom2025-ah, Rismani2024-md}.
Measures are necessary to quantify and monitor key \textit{attributes} -- observable properties of a system or a component -- and assess whether they meet predefined safety criteria \cite{Sanchez2010-vm}. 
From this perspective, a \textit{hazard} is a condition where one or more attributes deviate from expected parameters, creating the potential for a harmful outcome \cite{Leveson2011-fo,Leveson2018-zz}. A \textit{criterion} is the threshold or standard used to judge whether an attribute is within safe bounds; violating a criterion can signal a hazardous state \cite{Firesmith2004-an}. If a system enters a hazardous state, it could lead to \textit{harm} defined as the ``adverse lived experiences resulting from a system’s deployment and operation in the world'' \cite[p. 723]{Shelby2023-to}.
Figure \ref{fig:analyticalframe} illustrates this relational chain. 

System safety also frames measurement as a feedback mechanism that signals whether system elements are functioning within acceptable bounds, allowing actors to intervene when conditions drift toward harm \cite{Leveson2011-fo}. 
As shown in Figure \ref{fig:controldiagram}, measures of dataset quality, model performance, interaction logs, and perception, act as feedback signals for higher-level components and actors. Dataset quality measures can help data teams refine or rebalance data; model performance measures can guide design choices; and interaction and perception-based evaluations can reveal user experiences, enabling improvements to system behavior and outcomes.

In this work, we leverage these system safety concepts to analyze existing measures designed to assess adherence to AI ethics principles. This lens allows us to assess the adequacy of current evaluation practices not only in terms of \textit{what} they measure but also \textit{how effectively} they help anticipate and mitigate harm across the entire sociotechnical system and its embedded context. 

\section{Methodology}

\begin{figure*}[ht]
  \centering
  \includegraphics[width=\textwidth]{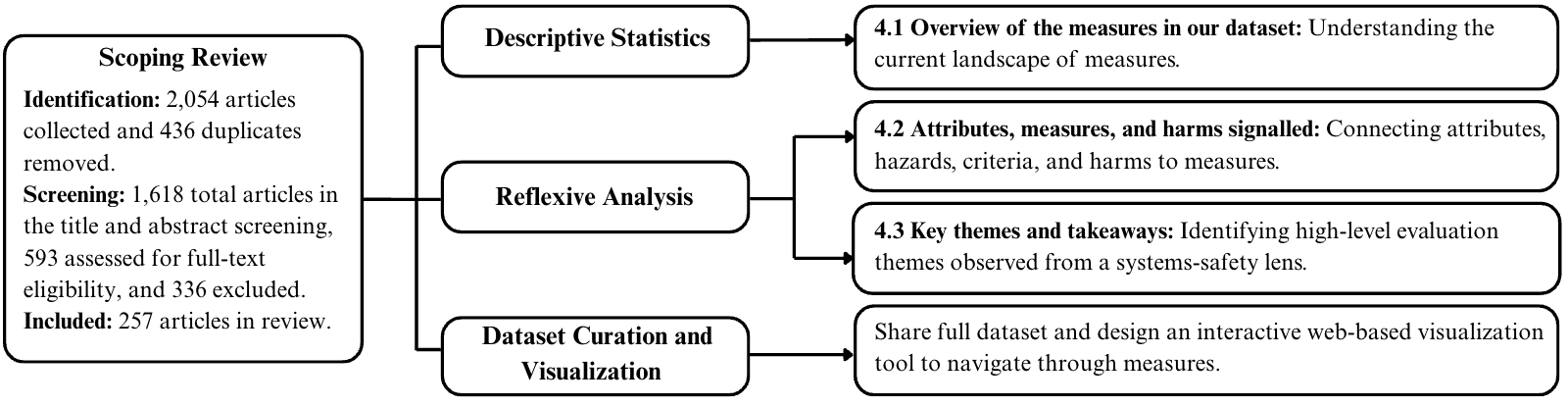}
  \caption{Overview of the primary methods conducted in this work.}
  \label{fig:method}
\end{figure*}

Our methodology, as illustrated in Figure \ref{fig:method} consisted of three steps: (1) a scoping review to identify measures from recent computing literature; (2) a reflexive analysis to map each measure to specific attributes, hazards, and harms; and (3) compiling the results into a dataset and interactive visualization to support further exploration. 

\subsection {Scoping Review: Identifying the Measures}
Our scoping review focused on measures related to the eleven AI ethics principles articulated by Jobin et al. (\citeyear{Jobin2019-kt}), given their comprehensive synthesis of ethical AI principles. We followed the five-stage scoping review process outlined by Arksey et al. (\citeyear{Arksey2005-qf}) given its suitability for mapping emerging fields (see also: \citet{Peters2021-ws}), using the Covidence platform to manage the review. 
The lead author oversaw the entire process, with other team members contributing to various stages.

\subsubsection{Identify research questions}
We posed the following foundational question to guide this scoping review: \textit{What measurements and measurement processes are computing researchers proposing to assess an AI system's adherence to ethical principles?} We focused on literature from computing-related venues, as technologists are the primary developers and implementers of these evaluation approaches. 

\subsubsection{Identify relevant studies}
We employed a three-phase search strategy combining database searches, citation-based reviews, and expert consultation. We queried the ACM Digital Library and IEEE Xplore via the Web of Science for papers published between January 2017 and August 2023. Our search queries incorporated terms related to measurement, AI, and the eleven principles stated by Jobin et al. (\citeyear{Jobin2019-kt}). Our initial search terms yielded over 8,000 results, warranting refinements. After testing 108 queries for relevance, we selected 22, which yielded 1,677 papers for screening. 
We then conducted a citation-based review on the papers selected for full-text review to identify 310 additional papers, including influential works from 2011–2016. Finally, three domain experts recommended an additional 40 papers for review. The collection of studies concluded in February 2024. 

\subsubsection{Study selection}
After removing 436 duplicates, 1,618 papers proceeded to title and abstract screening by four researchers using predefined inclusion and exclusion criteria. We included any peer-reviewed studies that proposed a measure to assess how a non-embodied AI system adheres to one of the 11 ethics principles from Jobin et al. (\citeyear{Jobin2019-kt}). Each paper was initially screened by at least two of the four researchers.  
Papers with two “yes” votes and two “no” votes were included or excluded, respectively. Conflicts were resolved by the lead author. This stage resulted in 593 papers being passed to full-text review, from which 257 papers were included for data extraction. A PRISMA flow diagram is available in Appendix A of the supplemental materials in our GitHub repository \cite{rismani2025measuring}.

\subsubsection{Data charting and summarization}

The data charting and summarization process involved two stages: extracting information from the papers and then separating multiple measures from each paper into distinct entries. A team of three researchers, including the lead author, performed this step. For each measure discussed in a paper, we used templates in Covidence and Google Sheets to extract key elements, including the ethics principle, measure name (e.g., demographic parity), measurement process (e.g., comparing positive classification rates), assessment type (e.g., statistical), the system component(s) being measured (e.g., output), the application area (e.g., criminal justice), and publication metadata. Drawing on the OECD's definition of an AI system \cite{OECD2024-ox}, we mapped each measure to one or more system components: data/input, model, output, user-output interaction, or the full system (as shown in Figure \ref{fig:controldiagram}).
In addition, we categorized measures into four assessment types: mathematical (based on formal equations or logic), statistical (derived from data distributions and patterns), behavioral (based on observed user actions), and self-reported (based on users’ subjective responses).  As many papers proposed multiple distinct measures, this process yielded a final corpus of 791 individual entries.

\subsection{Reflexive Analysis: Identifying the Criterion, Attribute, Hazard, and Harm}

For each of the 791 measures, we analyzed the measurement process taking place and the context provided in the source paper. We revisited source papers as needed to extract additional information. For each measure, we identified whether the author(s) provided an explicit criterion (e.g., equal positive prediction rates across groups) for the measure (e.g., demographic parity), and subsequently categorized the measure’s attribute (e.g., distribution of system output), hazard (e.g., disparate distribution of output), and related primary and secondary sociotechnical harms (e.g., allocative harm through the denial of loans or representational harm through the reinforcement of stereotypes). 

For criteria, attributes, and hazards, we developed inductive codes from our detailed review of the papers. For harm categorization, we applied the five categories from Shelby et al.'s sociotechnical harm taxonomy: allocative, representational, quality of service, interpersonal, and social system harms. Two researchers independently coded the first 100 measures and met to establish consensus. The remaining measures were coded collaboratively, with the lead author handling the majority of entries and consulting with the second researcher as needed to refine coding decisions. The full team then iteratively reviewed the completed dataset to identify key patterns connecting principles, attributes, hazards, and harms, using reflexive thematic analysis \cite{Braun_Clarke_2006,Braun_Clarke_2021}. 

\subsection{Dataset Compilation and Visualization}
The final stage involved data pre-processing, statistical analysis, and visualization. After cleaning the raw data for lexical inconsistencies, we calculated the counts and proportions of the measures across the 11 ethical principles, system components, and harm categories. 
The final dataset includes 18 columns covering all 791 measures, their corresponding measurement process, system component(s), assessment factors (criterion, attribute, and hazard), primary harm, secondary harm (if present), and publication metadata. 
To facilitate further exploration of this dataset, we created an interactive Sunburst diagram. The description of all the features, the full dataset, and the visualization are included in our GitHub repository \cite{rismani2025measuring}.

\subsection {Limitations of Our Study}
We acknowledge several limitations. First, our scoping review captures a snapshot of a rapidly evolving field; many relevant measures may be unpublished or exist only in preprints.
In addition, not all RAI measures are explicitly framed in relation to the AI ethics principles outlined by Jobin et al. (\citeyear{Jobin2019-kt}). Second, our analysis is shaped by our team's interpretive lens; it is inherently reflexive and introduces some subjectivity in selection, extraction, and categorization. Third, while we adopted a specific harm taxonomy, alternative frameworks exist that could offer complementary perspectives. 

\section{Findings}
We begin with an overview of the measures in our dataset. Next, we examine how these measures relate to harms by analyzing the attributes they quantify and the hazards they represent. We close by discussing key themes from our reflexive analysis.

\subsection{Overview of the Measures}
In this section, we provide a high-level overview of how the measures identified in our corpus (\textit{n} = 791) map onto AI ethics principles, components of AI systems, and types of sociotechnical harms. We observe clear patterns in which principles receive the most attention, which system components are most frequently measured, and how different categories of harms are represented. Notably, the sum of the calculated percentages exceeds 100\% as measures can be coded under multiple principles/components. Aggregate percentage ratios use 791 as the denominator, whereas component or harm-type breakdowns use totals within each principle.

\begin{figure}[t]
  \centering
  \includegraphics[width=\columnwidth]{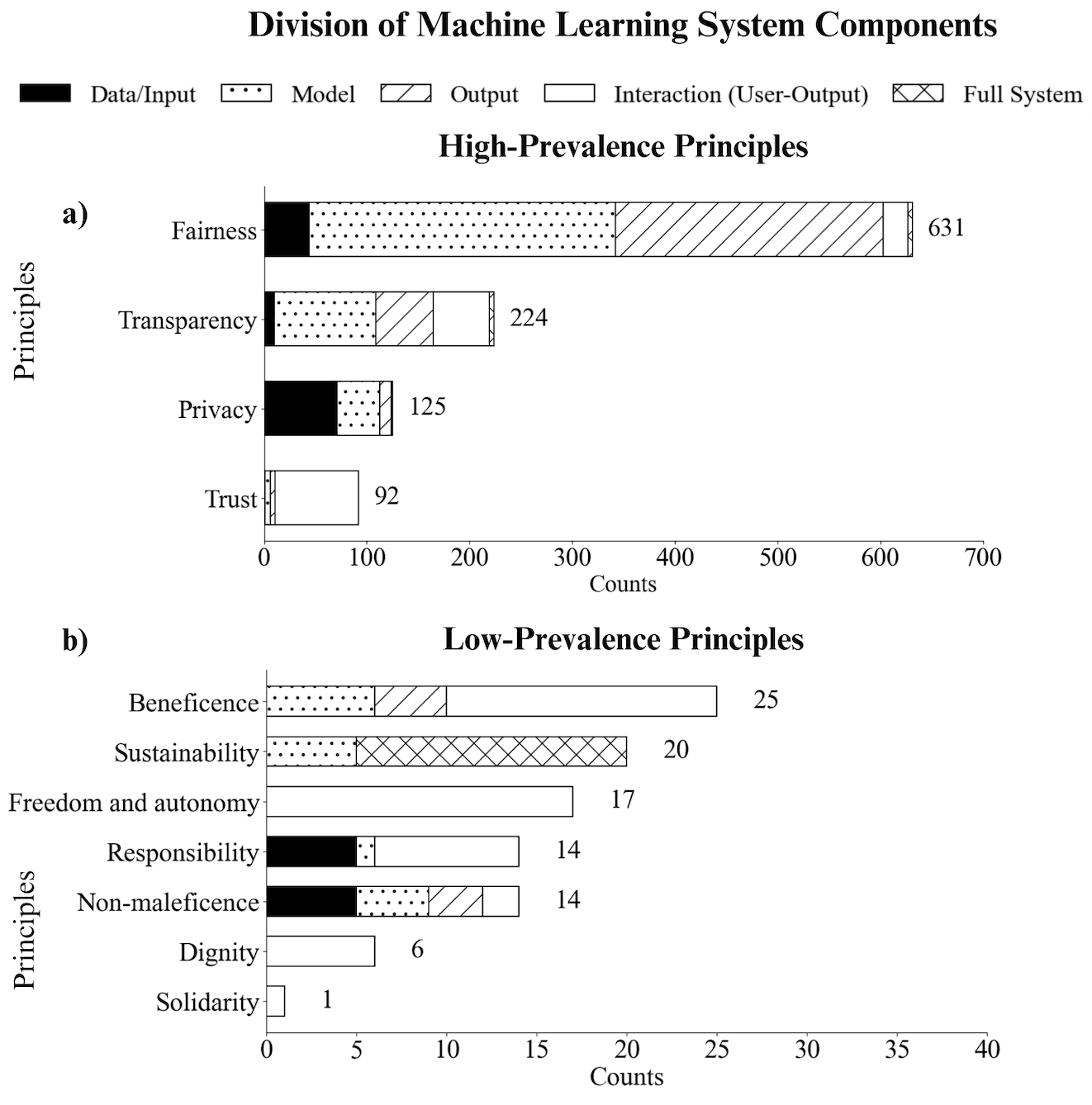}
  \caption{\small Division of machine learning system components categorized by (a) high-prevalence principles and (b) low-prevalence principles. The total count is above \textit{n = 791} as many measures are cross-listed with more than one principle and system component.}
  \label{fig:componentprinciple}
\end{figure}

\paragraph{Four principles account for 90\% of all measures}
We found a highly uneven distribution of measurement efforts across principles. Approximately 90\% (\textit{n} = 717) of measures coded fall under just four principles:  fairness (46.4\%, \textit{n} = 367), transparency (21.1\%, \textit{n} = 167), privacy (14.2\%, \textit{n} = 112), and trust (10.6\%, \textit{n} = 84), as shown in Figure \ref{fig:componentprinciple}. We refer to these as \textit{high-prevalence principles}. 
The remaining seven principles –- freedom and autonomy, beneficence, non-maleficence, solidarity, dignity, sustainability, and responsibility –- account for approximately 10\% (\textit{n} = 79) of the measures, which we term as \textit{low-prevalence principles}. Notably, a small subset of measures (2.9\%, \textit{n} = 23) were coded with more than one principle, such as transparency/trust, fairness/privacy, and dignity/fairness.

\paragraph{Disproportionate focus on model and output components}
We also observed a pronounced imbalance in which AI system components are evaluated. Across all measures, the vast majority target either the model (57.3\%, \textit{n} = 453) or the output (42.6\%, \textit{n} = 337). Far fewer assess the data/input (16.2\%, \textit{n} = 128) or the user-output interaction (25.4\%, \textit{n} = 201). 
This concentration is especially evident within the high-prevalence principles. For instance, 87.5\% (\textit{n} = 321) of all fairness-related measures assess either the model or the output. Similarly, 67.7\% (\textit{n} = 113) of transparency measures focus on the model or output, though a notable portion (32.3\%, \textit{n} = 54) also targets the user-output interaction. In contrast, privacy measures primarily assess attributes of the data/input (63.4\%, \textit{n} = 71), whereas trust measures overwhelmingly target the user–output interaction (96.4\%, \textit{n} = 81).  Among the low-prevalence principles, the majority of measures relate to the user-output interaction (62.0\%, \textit{n} = 49). Specifically, all measures related to the principles of solidarity, dignity, and freedom and autonomy are measures of user-output interaction. A key exception is sustainability, where 88.2\% (\textit{n} = 15) of measures assess the full system (e.g., total energy consumption \cite{henderson_towards_2020}), accounting for most of the full-system measures in our corpus.  


\paragraph{Uneven representation of harm types across principles} 
Our analysis reveals that existing measures span all five types of sociotechnical harm outlined by Shelby et al. (\citeyear{Shelby2023-to}), but their distribution is uneven. The most frequently addressed harms are interpersonal (29.2\%, \textit{n} = 231), involving instances where AI systems ``adversely shape relations between people or communities'' \cite[p. 727]{Shelby2023-to} and social system (25.9\%, \textit{n} = 205), reflecting ``adverse
macro-level effects'' of AI systems (p. 731). Allocative harms account for 25.0\% (\textit{n} = 198), capturing cases of economic and/or opportunity loss due to AI deployment. The remaining  measures correspond to representational harm (13.1\%, \textit{n} = 104) describing the reproduction of ``unjust societal hierarchies'', and quality of service harms (6.7\%, \textit{n} = 53) which reflect performance disparities across identity groups.

\begin{figure}[ht]
  \centering
  \includegraphics[width=\columnwidth]{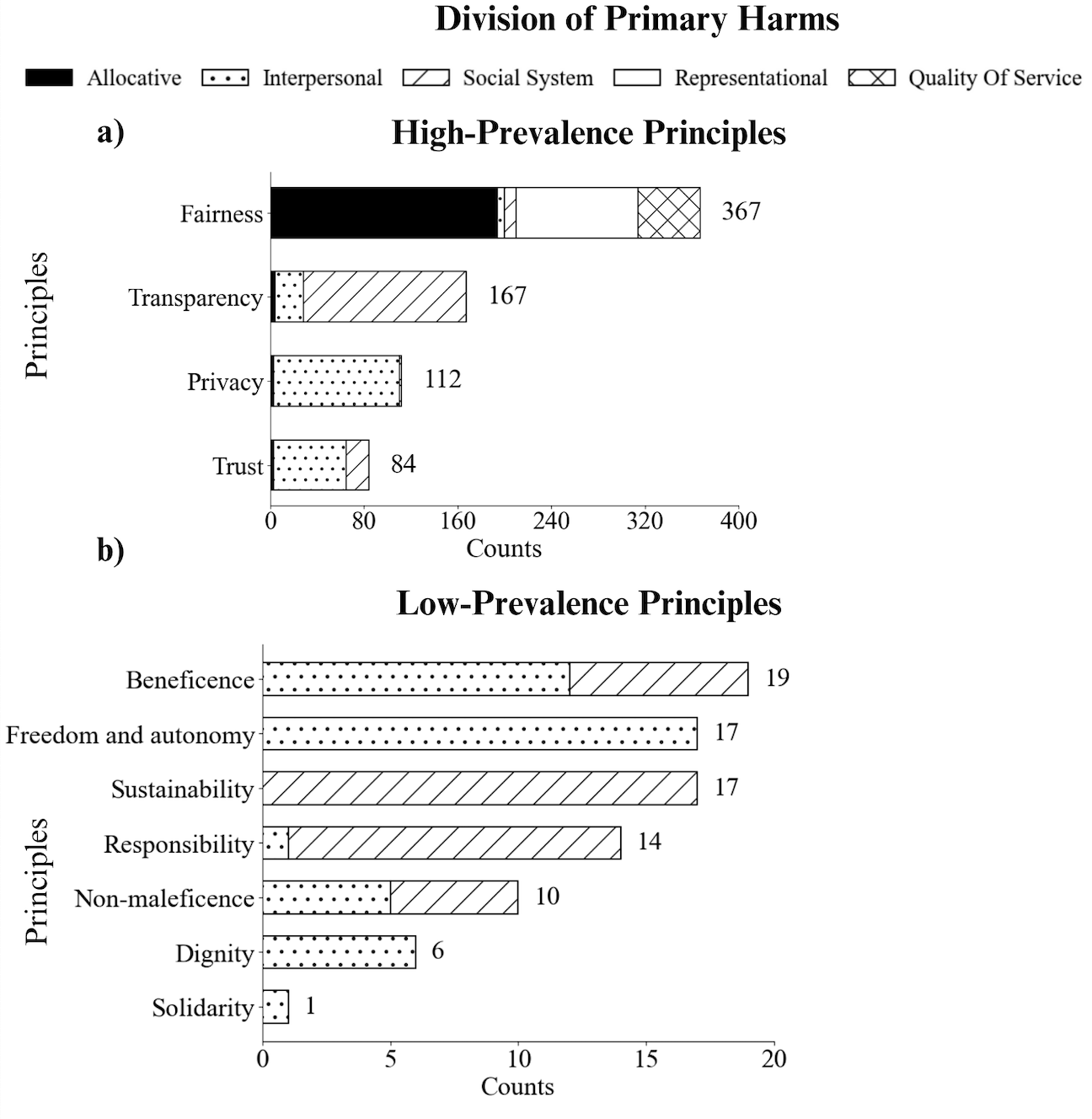}
  \caption{\small Division of primary harms categorized by (a) high-prevalence principles and (b) low-prevalence principles. }
  \label{fig:harmprinciple}
\end{figure}

Measures for each high-prevalence principle tend to concentrate on a few harm types (Figure 5). For example, 95.6\% (\textit{n} = 351) of fairness measures correspond to representational, allocative, or quality of service harms. The majority of transparency measures (83.2\%, \textit{n} = 139)  map to social system harms. Privacy measures are almost exclusively linked to interpersonal harms (95.5\%, \textit{n} = 107), while nearly all of the trust measures (96.4\%, \textit{n} = 81) are linked to interpersonal and social system harms. In contrast, measures tied to low-prevalence principles are only linked to interpersonal (53.2\%, \textit{n} = 42) or social system harms (46.8\%, \textit{n} = 37).

\subsection{Attributes, Hazards, and Harms Signaled}
In this section, we examine the attributes of AI systems quantified by the measures and the hazards they signal for each of the five harm types. The total number of measures related to each harm type is presented in parentheses along the section headings. For each type of harm, we identify and describe the most common categories of measures, system components, attributes, and associated hazards. 
Note that many of the measures in our corpus can be interpreted through the lens of multiple types of harms. In such cases, we categorized the measures based on the original framing of the measures in the source articles. In addition, common hazards include disparities across system outcomes, errors, and performance between groups, which occur across representational, allocative, and quality-of-service harm types. While these hazards often involve similar measures, we assigned harm types according to the framing in the source paper. A more detailed and illustrative thematic summary of each of these mappings can be found in Appendix C in the supplementary materials on our GitHub \cite{rismani2025measuring}.

\subsubsection{Representational harm (104 measures)}

Representational harms ``occur when algorithmic systems reinforce the subordination of social groups along the lines of identity" (Shelby, \citeyear{Shelby2023-to}, p. 728, as cited in \citeauthor{Katzman2021-rq}, \citeyear{Katzman2021-rq}). 
Most measures signaling this harm are mathematical or statistical assessments, quantifying attributes of the model or output (79.8\% and 71.2\%, respectively). A smaller share (19.2\%) of measures target the data/input component, and none relate to the full system or user-output interaction stages. 
We identified three categories of measures signaling representational harm which we described below.

\textbf{Representation of identities:} These measures quantify how different social groups are represented in the data/input, model, and output of an AI system. 
Measures associated with the data/input component most often point to the hazard of \textit{imbalanced group representation in the training data.}
For instance, \textit{person prominence} is a measure of how prominently individuals from different groups appear in visual training content \cite{Wang_2020}. Additionally, the \textit{difference in reconstruction error} captures disparities in how accurately different groups are represented in compressed data \cite{Pelegrina_2022}. 
The measures that observe attributes of a model signal the \textit{reinforcement of stereotypes through learned embeddings} as a hazard. They typically quantify how different groups are represented within the model.
Included in this category are measures like the \textit{cosine similarity of images} and \textit{word embedding association tests}, which quantify stereotypical associations by analyzing patterns in the model embedding space \cite{Goyal_2022,Malik_2022}. The measures that observe the output of an AI system quantify the presence of protected groups in its results. They signal an \textit{underrepresentation or omission of specific protected groups in the outcomes of an AI system} as a hazard. Examples include the \textit{diversity score} measure, which reflects the variety of representation across social attributes in the system's outcomes, and the \textit{percentage of content by group} measure, which quantifies the proportion of representation of specific identity groups \cite{Solaiman2023-ol,Ferraro_2021}.   

\textbf{Reinforcement of social hierarchies:} These measures focus on assessing who receives disproportionate errors or benefits, directing attention to whether the system’s outcomes are distributed in ways that reinforce existing social hierarchies. 
The measures and hazards are similar to those associated with allocative harm, which are further described in the following section under the category of \textit{disparate outcome distribution}.


\textbf{Misalignment with user identity:} These measures assess the degree to which system outputs resonate with users' identities, values, and cultural contexts.
For example, the \textit{steerability} measure quantifies how well an AI model can align its responses with the perspectives of specific demographic groups when prompted with contextual cues \cite{santurkar_whose_2023}, and the \textit{commonality} measure indicates whether users collectively encounter content that fosters a shared cultural experience \cite{Ferraro_2024}. The key hazard associated with this perspective is that an AI system \textit{creates outcomes that are irrelevant, unfamiliar, or misaligned with users’ identities and experiences}.


\subsubsection{Allocative harm (198 measures)}
A hazard enables allocative harm when a system's distribution of information, resources, or opportunities adversely affects historically marginalized groups (Shelby, \citeyear{Shelby2023-to}, p. 729, as cited in \citeauthor{barocas-hardt-narayanan}, \citeyear{barocas-hardt-narayanan} and \citeauthor{Richardson_Gilbert_2021}, \citeyear{Richardson_Gilbert_2021}).
The vast majority of measures associated with this harm type quantify attributes of the model (83.8\%) and its output (69.7\%) using statistical or mathematical approaches. A smaller proportion observes attributes of the data/input (11.6\%). Measures of user-output interaction account for 7.1\% of the total. 
Compared to representational harm, this set contains a higher proportion of self-reported and behavioral measures. Overall, we identified four categories of allocative harm measures as described below.

\textbf{Disparate outcome distribution:} These measures quantify how beneficial or erroneous system outputs are distributed across different social groups.
For example, \textit{disparate impact} measures the ratio of positive outcomes (e.g., loan approvals) between disadvantaged and advantaged groups \cite{Besse_2021}. \textit{Individual-user-to-individual-item fairness} indicates whether each user–item pair receives exposure aligned with a fairness target \cite{Wu_2018}.
Both measures signal the hazard that \textit{certain groups systematically receive fewer beneficial outcomes from the AI system}. 
Similarly, measures of error distribution, such as \textit{treatment equality} (the ratio of false negatives to false positives) \cite{berk_fairness_2021}, help observe how model errors are distributed across groups. 
These measures signal the hazard that \textit{error rates disproportionately affect specific groups}.

\textbf{Disparate model performance:} These measures capture
\textit{group-level differences in model performance}, leading to disparities in access to resources or opportunities. The measures and hazards are similar to those associated with the quality of service harms described under the category of \textit{performance quality disparities} in the following section.

\textbf{Perceived decision fairness:} These measures
capture users' fairness perception of AI systems. These are primarily measures of user-output interaction, which quantify a user's perceived understanding of algorithmic decisions 
and perceived fairness of the AI-based process taking place. Examples include self-reported understanding (e.g., agreement with “I understand the process by which the decision was made”) \cite{Binns_2018} and perceived fairness ratings after viewing explanations of a training dataset \cite{Anik_2021}. These measures signal the hazard that \textit{users may not understand an AI system’s reasoning or may perceive its decisions to be biased or unjust.}

\textbf{Design-level disparities:} These measures 
capture upstream design choices that lead to outcome disparities, such as the decisions surrounding features and labels in the input dataset and model design choices that incorporate fairness considerations. 
Examples include \textit{class imbalance and conditional disparity in labels} measures, which quantify gaps in representation or label distribution across groups in a dataset \cite{Hardt_2021}, and the \textit{price of fairness} measure, which captures the accuracy-fairness trade-off by quantifying the increase in model loss required to reduce disparity \cite{berk_convex_2017}. 
Both measures signal the hazard that \textit{poor quality data or modeling strategies may produce or reinforce disparities in system outcomes.}

\subsubsection{Quality of service harm (53 measures)}

Quality of service harms ``occur when algorithmic systems disproportionately underperform for certain groups of people along social categories.'' (Shelby, \citeyear{Shelby2023-to}, p. 730). 
Most of the measures (90.6\%) signaling this harm type assess the model or output components of an AI system using statistical or mathematical methods. Only five measures are classified at the user-output interaction, and none are found for the data/input or full system components. 
We observed two categories of measures, which we describe next.

\textbf{Performance quality disparities:} These measures capture
differences in model performance across groups using measures like \textit{precision, F1-score,} or \textit{exposure}. Whereas allocative harm focuses on unequal access to resources, these measures focus on whether the system functions reliably and accurately for all users, signaling the hazard that \textit{the model performs unevenly across certain groups}.


\textbf{User experience quality:} These measures capture 
the perceived quality of the user-output interaction among a user(s) and an AI system.
For instance, the \textit{Hellinger distance} measure can quantify the divergence between the distribution of recommended content across demographics (e.g., gender of artists) and a user's historical behavior \cite{Ferraro_2021}. A high Hellinger distance indicates that recommendations deviate significantly from a user's past preferences, signaling potential misalignment. The resulting hazard is that \textit{system outputs are less useful and satisfactory for certain groups.}

\subsubsection{Interpersonal harm (231 measures)}

Interpersonal harm refers to ``instances when algorithmic systems adversely shape relations between people or communities" \cite[p. 730]{Shelby2023-to}. 
The associated measures provide observations of privacy violations, loss of agency, and diminished well-being. Notably, we did not identify any measures for the technology-facilitated violence harm sub-type. Overall, the measures in this harm category assess the user-output interaction (49.4\%), data/input (31.2\%), and model (22.5\%) components, with a few measures relating to the output (7.4\%). Among this harm type, measurement approaches are diverse, with 50.2\% being mathematical or statistical, 38.5\%  self-reported, and 12.1\%  behavioral. We identified five different categories, which we describe next. 

\textbf{Design-level privacy preservation:} 
These measures indicate the risk of re-identification or the level of privacy protection applied to a dataset and/or model. 
Example measures include the \textit{inverse trace of the Fisher Information Matrix}, which quantifies the difficulty of reconstructing sensitive variables from released data \cite{Farokhi_2021}; the \textit{K-value}, representing the minimum group size required to make individual records indistinguishable \cite{Gonz_lez_Zelaya_2023}; and the differential privacy parameters $\epsilon$ and $\delta$, which quantify the maximum privacy loss and the probability that this guarantee fails \cite{Jarin_2022}. These measures signal the hazards that \textit{individuals can be identified due to patterns in the dataset} or \textit{privacy protections are either too weak to be effective or so strong that they undermine data utility.}

\textbf{Privacy attack risk:} These measures reflect 
the model's susceptibility to adversarial inference and the amount of sensitive information that can be extracted. For instance, the \textit{adversary accuracy} measure estimates the likelihood that an attacker correctly infers a sensitive label \cite{Xiao_2019}, while \textit{mutual information} measures how much private information is leaked through model outputs \cite{Zhang_2018}. These measures relate to the hazard of \textit{successful privacy attacks that expose user-level information} and \textit{inference methods that allow significant reconstruction of individual data.}

\textbf{Influence on user behavior:} This category of measures captures the AI system's influence on the user. For example, \textit{acceptance rates} quantify how often users follow AI-generated suggestions \cite{Zhang_2023}, and \textit{decision threshold shifts} measure how much a user's decision boundary changes when aided by the AI system \cite{Poursabzi_Sangdeh_2021}. These signal the hazard that users may be \textit{unduly influenced or become over-reliant on AI system outputs}.

\textbf{Perceived level of influence}: Related to the previous category, these measures quantify
users' perceived control and perceived autonomy of the AI system. \textit{Perceived control Likert scales}, for example, capture a users' sense of influence over outcomes, while \textit{mind perception agency scales} reflect how much autonomy and independent decision-making users attribute to the AI system \cite{Kim_2023,Gonzalez_2022}. These measures signal the hazard that \textit{users feel disempowered or incorrectly believe the system acts with more autonomy than it actually possesses.}

\textbf{Impact on user well-being:} These measures observe the impact of a system's output on user well-being, 
quantifying attributes like the harmfulness of generated content, perceived quality of the system output, and perceived respectfulness. The \textit{HONEST score}, computed as the proportion of completions containing offensive content, is an example measure of the harmfulness of generated content \cite{Nozza_2021}. 
\textit{Perceived helpfulness and risk ratings} capture users' subjective impressions of an AI system's utility or danger \cite{Przyby_a_2021}, 
and \textit{dehumanization scales} offer a self-reported measure of whether users feel objectified or treated as less than fully human in their interactions with AI \cite{Formosa_2022}. These measures signal hazards such as an \textit{exposure to harmful content}, \textit{distorted perceptions of system utility or risk}, and a \textit{lack of perceived dignity or respect in AI-mediated experiences}.

\subsubsection{Social system harm (205 measures)}
Social system harms ``reflect the adverse macro-level effects of new and reconfigurable algorithmic systems, such as systematizing bias and inequality and accelerating the scale of harm.'' (Shelby \citeyear{Shelby2023-to}, p. 731)
All measures in this category relate to information loss or environmental harms; none reflect the harm sub-types of cultural, socio-economic, and political/civic harm. The measures are distributed across multiple system components: model (52.2\%), output (31.2\%), user–output interaction (33.2\%), full system (10.7\%), and data/input (6.3\%). About three-quarters are mathematical or statistical; the remainder are behavioral or self-reported assessments. We identified two main categories as described below.

\textbf{Understanding of the system}: These measures quantify 
the interpretability and understandability of the system, and the quality of the post-hoc method used to explain model predictions. 
The \textit{perceived understanding} measure, for example, captures users' self-reported comprehension and ability to anticipate the model's behavior \cite{Wang_2022}. 
An example of post-hoc method quality measures includes the \textit{fidelity score}, which reflects how much a model's performance drops when features identified as influential are masked \cite{Bonet_2022}. \textit{Explanation completeness} is another example, which calculates the fraction of the overall explanation retained in a reduced subset \cite{Mariotti_2022}. These measures signal hazards that \textit{users are unable to meaningfully interpret the system} or \textit{are misled by low-fidelity explanations}, raising the risk that AI systems obscure underlying logic, or contribute to misinformed decision-making.

\textbf{Resource intensity}: These measures signal the hazard that \textit{AI systems consume excessive resources and generate high emissions}.
Energy consumption and carbon emissions are two attributes observed in these measures. For example, \textit{average energy consumption} can be estimated from hardware-specific multiply-and-accumulate operations, while the \textit{energy per sample} measure is calculated by dividing the real-time power usage by the model's processing throughput \cite{Yang_2017}. Carbon impact is captured through measures such as the \textit{estimated greenhouse gas emissions}, which factors in regional electricity carbon intensity, and the \textit{CO\textsubscript{2}-equivalent output}, calculated using tools like the ML Emissions Calculator based on training time, hardware type, and location \cite{Savazzi_2023}.

\subsection{Key Themes and Takeaways}
Our analysis reveals five cross-cutting themes about the current state of RAI measurement.

\subsubsection{Measures are fragmented across system components}
Most evaluation measures target isolated components of AI systems, particularly the model or its outputs, while overlooking how harms emerge from interactions across components. This lack of system-spanning evaluation has been noted before. For example, Black et al. (\citeyear{Black2023-cc}) highlight that most fairness interventions focus on the modeling stage, leaving earlier and later stages under-examined. A fairness measure of a dataset does not equate to the fairness performance of an AI system trained on it \cite{Barocas2021-sr}. Likewise, a high-performing AI model that leaves users feeling unduly influenced can lead to unsafe AI system use.

Notably,  a small subset of measures in our corpus explicitly engage with component interdependencies or aim to evaluate the system holistically. One example is \textit{MARS-Gym}, a benchmark framework for reinforcement learning-based recommender systems that incorporates fairness measures, such as disparate treatment and mistreatment, to trace how sensitive inputs influence model decisions and affect outputs. Its design enables evaluation across the data and input, model, and outcome components, offering a more interconnected view of system behavior \cite{Santana_2020}. A few sustainability-related measures, such as those by Savazzi et al. (\citeyear{Savazzi_2023}) and Ollivier et al. (\citeyear{Ollivier_2023}), provide observations about the full system. These measures estimate environmental impact by combining energy consumption across various stages of development. While they do not model inter-component effects directly, they integrate attributes across stages to reflect overall system performance. To identify harms at the systems-level, assessments must go beyond modular checkpoints to examine how one component’s attributes affect another, capturing dependencies and dynamics that could create unsafe outcomes.


\subsubsection{Measurements are taken far from where harm is experienced}
There is a large gap between where normative measurements are taken and where harms are ultimately experienced. While many papers in our corpus are motivated to prevent potential sociotechnical harms in a general sense, the process of quantifying the attributes into measures or taking measurements from components of a system is often decoupled from a specific harm and the conditions under which it occurs. Critically, we found no instances where a component-level measurement was empirically traced to a downstream harm experienced by users. This disconnect reflects criticisms raised about the validity of common measures  \cite{Jacobs2021-bj,Blodgett2021-fg}.

In our corpus, many measures are designed to evaluate system-level hazards instead of direct harm assessments. For example, the \textit{HONEST score} measure quantifies the presence of offensive language in model outputs \cite{Nozza_2021}. While it observes a user-facing attribute, it does not directly assess the lived experience of the users. Similarly, Shen et al.'s (\citeyear{Shen_2023}) \textit{correlated sensitivity metric} estimates privacy leak risk based on dataset composition . While it highlights a potential privacy hazard, using it as a measure of harm is questionable, as it does not assess the experience of privacy violation. Crucially, the disconnect lies not in the measures themselves, but in how they are applied and interpreted within evaluation practices. As Barocas et al. (\citeyear{Barocas2021-sr}) argue, component-level measures can help identify hazards but offer limited insight into the lived experience of harm. We argue that evaluation practices should distinguish between measures that assess hazards and those that evaluate harm as lived experiences. The mapping presented in this paper offers an initial step toward this goal.

\subsubsection{Measures lack clear criteria for identifying hazards}

Only a small subset of measures -- approximately 26.9\% (\textit{n} = 213) -- were accompanied by explicit criteria for interpretation. While not every measure requires a predefined criterion, such guidance becomes important when a measure is used to identify potential hazards or signal when system behavior may be unacceptable. From the measures that had stated criteria, few provided a discussion on appropriate thresholds. This is not surprising as thresholds are inherently context-dependent and may vary based on application domains, user populations, or risk tolerances. 
Most of the measures with stated criteria relate to fairness and focus on concepts such as group- or individual-level parity (e.g. \textit{demographic parity criterion ensures that individuals with a protected attribute have the same likelihood of being classified as positive (e.g., recidivist) as those in a reference group}, as paraphrased from \cite{Tolan_2019}). However, achieving parity is rarely feasible or expected. As a result, practitioners rely on judgment calls or heuristics such as the disparate impact rule, which states that the selection rate for a disadvantaged group should be at least 80\% of that for the most advantaged group. 
Threshold ambiguity was also present in transparency measures. Example criteria include \textit{good explanations should be sparse}  or \textit{good explanations should be stable}, as paraphrased from\cite{Yuan_2022}. Yet these criteria are under-specified for operationalization. For example, determining how much detail is needed in a feature description or what level of consistency qualifies as stable are inherently subjective decisions, often left to the discretion of practitioners. 
From a systems safety perspective, thresholds are essential for distinguishing tolerable variation from conditions that signal a hazard. Evaluation practices can greatly benefit from outlining and disclosing what is deemed to be unacceptable behavior of a system component or a full system in relation to an attribute that is being measured. This clarity enables both those designing evaluations and those applying them to make informed, safety-aligned decisions.

\subsubsection{Measures overlook how harms accumulate over time}

Although sociotechnical harms stemming from AI systems often emerge gradually, many of the measures are taken at a single point in time. While some statistical measures related to models and outputs are integrated into development and maintenance pipelines, many are not applied over time. Notably, none of the papers in our corpus discussed how their proposed measures could support long-term monitoring. A few papers related to environmental harm were exceptions, tracking energy consumption or carbon emissions throughout training and deployment \cite{henderson_towards_2020}. These were among the only examples in our dataset where time was meaningfully integrated into the construction of the measure itself. From a system safety perspective, time-aware evaluation is critical for detecting emergent hazards and triggering timely intervention \cite{Leveson2018-zz}. Studying how component-level attributes change over time, and when those changes signal a hazard and harm, is a valuable direction for future evaluation practice. 

\subsubsection{The role of perception-based measures in signaling harms remains unclear}
Perception-based measures composed 17.3\% (\textit{n} = 137) of the entire corpus. They are particularly prevalent in measures of interpersonal and social-system harms. These self-reported measures are central to evaluating user experiences. However, in our reflexive analysis, it was difficult to determine what hazards and harms these measures signal. Human perception is subjective, and self-reported measures have well-established drawbacks \cite{Lai2023-mo}. Therefore, it is not surprising when users’ perceptions of a system (e.g., perceived utility) do not align with other measures (e.g., actual system efficacy).
While optimizing for a perception-based measure is inappropriate for designing safe AI, ignoring these measures could mean overlooking potential indications of harm experienced by users. 
The RAI community needs to pay more attention to unpacking the role perception-based measures can play in harm and risk assessments of AI systems. 


\section{Discussion}
\subsection{Implications for Governance and Policy}

Measures are central to emerging AI governance and evaluation frameworks, whether they are principle-based, risk-based, or harm-based \cite{Jacobs2021-ts, Brundage2020-ck}. Our analysis highlights several critical gaps. First, existing measures disproportionately focus on a narrow set of principles (recorded as high-prevalence principles). Second, measures tend to concentrate on a limited set of well-established hazards which risks narrowing the scope of what gets measured and prioritized. This raises a deeper question: what hazards should we actually be paying attention to when it comes to understanding and mitigating harm? Without reflection and consensus on which hazards matter across contexts, important dimensions of harm may be overlooked \cite{Dobbe2024-tk, Selbst2019-kz}. This resulting fragmentation is further compounded by the lack of clear evaluation criteria for which levels of measurement indicate the presence or severity of a hazard.
For normative quality assessments of AI systems (e.g., AI ethics audits) to become a meaningful part of AI governance frameworks, a more comprehensive approach to identifying and monitoring harms and assessing adherence to a variety of ethical principles is needed \cite{Ojewale2025-uf}. By making explicit the connections between what is being measured, why it matters, and how it relates to potential harms, evaluation practices can become more traceable and accountable \cite{Kroll2021-nw}. 
Our corpus of measures and the accompanying visualization can serve as a practical and reflexive resource to springboard such an effort.

\subsection{Implications for Industry and Practice}

Our analysis offers two key implications for AI practitioners. First, our mapping of measures to principles, system components, assessment types, hazards, and harms can help practitioners select or design measures in a more deliberate and structured way. 
Rather than creating ad hoc measures, teams can use the visualization of our dataset to explore a variety of measures aligned with their evaluation goals, system attributes, or regulatory needs. Second, with an understanding that measurement selection, application, and implementation can be resource-intensive, this work's framing supports more strategic prioritization \cite{Madaio2022-vt}. By making explicit what a given measure captures-- and what it does not-- teams can better weigh trade-offs, justify their evaluation choices, and understand why one measure may be more appropriate than another \cite{Barocas2021-sr}. This can improve internal coordination across product, compliance, and management roles while making evaluation efforts more transparent and defensible \cite{Balayn2023-qo}. 

\subsection{Implications for Researchers}

We find two key implications for the research community. First, there is a pressing need for more structured and transparent reporting when new measures are introduced. Many papers lack clarity on what component is assessed, which attribute is measured, or how the measure relates to a specific harm or hazard. Without this scaffolding, measures are difficult to interpret or apply beyond their original context \cite{Olteanu2025-vq}.  Second, addressing the evaluation gaps identified (i.e., a disproportionate emphasis on certain system components and a narrow concentration on well-established hazards) requires stronger interdisciplinary collaboration \cite{Kawakami2025AIMeasurement}. Limitations such as narrow principle coverage, limited attention to cross-component interactions, and the disconnect between where measures are applied and where harm is experienced highlight the need to move beyond computing-centric approaches. 
Engaging with fields such as psychology, media studies, and sociology can help develop measures that better reflect user experiences, integrate perception-based data more meaningfully, and bring systems-level behavior into closer alignment with the experiences of harm. 

\subsection{Future Work}
Looking ahead, there are several promising directions for future work. While our analysis focused on application-specific systems, the growing use of general-purpose AI models introduces a new layer of complexity. These systems are harder to assess, with broader and less predictable use cases. They require more adaptive, multi-layered approaches to evaluation \cite{Wallach2024-xb, Weidinger2023-pe}. In addition, many organizations contribute to the development and deployment of different components within a product that uses general-purpose AI systems. Applying a system safety perspective to these evaluations can illuminate which system components are being assessed, where measures are missing, and which organizations are responsible for each part. Future work could also encourage more community reporting on how measures are used in practice across different types of AI systems \cite{Dev2021-mg}. In parallel, we see value in expanding how we conceptualize system components, beyond the five we used, to better reflect the complexity of real-world deployments. Finally, the visualization tool we developed could be refined and tested for use within organizations. Its usefulness will likely vary across sectors, company sizes, and governance structures. These directions point to a broader shift: treating evaluation not as an isolated research task but as a core part of the sociotechnical infrastructure needed to support RAI.

\section{Conclusion}
This paper offers a systems-level analysis of how AI measures relate to ethical principles, system components, and sociotechnical harms. By applying a systems safety lens, we find that current practices lack measurements that capture cross-component dynamics and that a significant gap remains between where measurements are taken and where real-world harms occur. 
Our analysis, dataset, and visualization tool aim to support more grounded, transparent, and context-sensitive evaluation practices. 
There is a pressing need for measurement approaches that reflect how harms emerge within sociotechnical systems. This work takes a step toward addressing this need by offering both a critical mapping of the current landscape and a foundation for building more comprehensive evaluation infrastructures.

\section*{Positionality Statement}
Our team, based in the United States and Canada, brings expertise from computer science, software engineering, HCI, robotics, sociology, and STS, with experience across industry and academia. This positioning shaped our approach to measurement, interpretation, and analysis. While we followed a rigorous process with multiple quality checks, we acknowledge the inherent subjectivity in our decisions, from selecting and extracting data to framing our analysis. Our familiarity with industry and academic measurement practices informed our understanding, particularly when resolving inconsistencies or gaps in the reviewed work. We aimed to critically engage with these measures while ensuring our interpretations remained faithful to the original sources.

\section*{Acknowledgments}
We gratefully acknowledge Ava Gilmor and Bonam Mingole for their assistance with the scoping review, and Fernando Diaz for his insightful feedback throughout this project. This project was financially supported by Google.

\bibliography{aaai25}

\end{document}